\documentclass[prb,preprint,aps,superscriptaddress,longbibliography]{revtex4-2}
\usepackage{epsfig}
\usepackage{amsmath}
\usepackage{amssymb}
\usepackage{threeparttable}
\usepackage{graphicx}
\usepackage{natbib}
\usepackage{longtable}
\usepackage{txfonts}
\usepackage{color}
\usepackage{braket}
\usepackage{revsymb}
\usepackage{amsfonts,amssymb}
\usepackage{xr}
\usepackage{verbatim}
\usepackage{algorithmic}
\usepackage{algorithm}
\usepackage{titlesec}
\usepackage{indentfirst}
\usepackage{svg}

\titleformat{\paragraph}
{\normalfont\normalsize\bfseries}{\theparagraph}{1em}{}
\titlespacing*{\paragraph}
{0pt}{3.25ex plus 1ex minus .2ex}{1.5ex plus .2ex}

\begin{document}
\makeatletter
\newenvironment{breakablealgorithm}
  {
   \begin{center}
     \refstepcounter{algorithm}
     \hrule height.8pt depth0pt \kern2pt
     \renewcommand{\caption}[2][\relax]{
       {\raggedright\textbf{\ALG@name~\thealgorithm} ##2\par}%
       \ifx\relax##1\relax 
         \addcontentsline{loa}{algorithm}{\protect\numberline{\thealgorithm}##2}%
       \else 
         \addcontentsline{loa}{algorithm}{\protect\numberline{\thealgorithm}##1}%
       \fi
       \kern2pt\hrule\kern2pt
     }
  }{
     \kern2pt\hrule\relax
   \end{center}
  }
\makeatother

\title{Discovery of Stable Hybrid Organic-inorganic Double Perovskites for High-performance Solar Cells via Machine-learning Algorithms and Crystal Graph Convolution Neural Network Method}

\author{Linkang Zhan}
\affiliation{WLSA Shanghai Academy, 2 Zhengxi Road, Yangpu District, Shanghai 200243, China}

\author{Danfeng Ye}
\affiliation{WLSA Shanghai Academy, 2 Zhengxi Road, Yangpu District, Shanghai 200243, China}

\author{Xinjian Qiu}
\affiliation{WLSA Shanghai Academy, 2 Zhengxi Road, Yangpu District, Shanghai 200243, China}

\author{Yan Cen}
\email{cenyan@fudan.edu.cn}
\affiliation{Department of Physics, Fudan University, No.220 Handan Road, Shanghai 200243, China}

\begin{abstract}
Hybrid peroskite solar cells are newly emergent high-performance photovoltaic devices, which suffer from disadvantages such as toxic elements, short-term stabilities, and so on. Searching for alternative perovskites with high photovoltaic performances and thermally stabilities is urgent in this field. In this work, stimulated by the recently proposed materials-genome initiative project, firstly we build classical machine-learning algorithms for the models of formation energies, bangdaps and Deybe temperatures for hybrid organic-inorganic double perovskites, then we choose the high-precision models to screen a large scale of double-perovskite chemical space, to filter out good pervoskite candidates for solar cells. We also analyze features of importances for the the three target properties to reveal the underlying mechanisms and discover the typical characteristics of high-performances double perovskites. Secondly we adopt the Crystal graph convolution neural network (CGCNN), to build precise model for bandgaps of perovskites for further filtering. Finally we use the \textit{ab-initio} method to verify the results predicted by the CGCNN method, and find that, six out of twenty randomly chosen (CH$_3$)$_2$NH$_2$-based HOIDP candidates possess finite bandgaps, and especially, (CH$_3$)$_2$NH$_2$AuSbCl$_6$ and (CH$_3$)$_2$NH$_2$CsPdF$_6$ possess the bandgaps of 0.633 eV and 0.504 eV, which are appropriate for photovoltaic applications. Our work not only provides a large scale of potential high-performance double-perovskite candidates for futural experimental or theoretical verification, but also showcases the effective and powerful prediction of the combined ML and CGCNN method proposed for the first time here.
\end{abstract}

\maketitle

\newpage
\tableofcontents

\section{Introduction}

Since the beginning of the new century, with the population explosion and rapid economic and social development, human beings have a great demand for energy, and industrial production has also caused great damage to the environment, resulting in the energy crisis, environmental crisis and climate crisis becoming the focus of attention in recent years, which also seriously affects the daily production and life of human beings. The search for sustainable new green energy sources to replace highly polluting fossil energy is one of the main solutions to cope with the above-mentioned crisis, among which photovoltaic energy derived from the photovoltaic effect is the focus of development. According to the global action plan formulated by the International Sustainable Energy Agency (IRENA)\cite{dhabi2020irena}, by 2050, the total installed capacity of photovoltaic power generation should reach 14 TW, while as of 2020, the global installed capacity is 0.7 TW, which is far below the target, and countries around the world urgently need to continue to vigorously develop photovoltaic power generation technology.

Therefore, since 1954, Bell Labs successfully manufactured the first solar cell and achieve 4.5\% energy conversion efficiency, photovoltaic cells through three generations of technology evolution\cite{fraas2014history}, the mainstream of the solar-cell market today is the first generation of monocrystalline or polycrystalline silicon-based solar cells and the second generation of thin film cells, but silicon-based photovoltaic solar cells suffer from high production costs, environmental pollution and other problems. So in recent years, the third generation of photovoltaic cells with high power conversion efficiency (PCE) has become a research hotspot. Among them, the organic-inorganic lead halid hybrid perovskites represented by CH$_3$NH$_3$PbI$_3$ (MAPbI$_3$) have attracted widespread attention\cite{zhao2016organic,kojima2009}. The hybrid organic-organic perovskites (HOIP) ABX$_3$ generally contain an A-site organic group, B-site is usually Pb or Sn, and X-site is a halogen element. Since its first application in photovoltaic cells in 2009, achieving a photovoltaic conversion efficiency of 3.8\%\cite{kojima2009}, after more than ten years of development, not only a wide range of types, its single junction PCE has also exceeded 25\%\cite{min2021perovskite}, comparable to the current mainstream silicon-based solar cells, and its commercialization process is also developing rapidly, and may be widely used in the near future.

Although it has the advantages of easy preparation, low cost, and good photovoltaic performance, MAPbI$_3$ has very obvious disadvantages: it contains toxic element Pb, has poor thermal conductivity, and does not work stably under the ambient environment\cite{rao2021review}. Therefore the search for its replacement with environmentally friendly and long-term stability is urgent in this field. Traditionally, the invention and promotion of new materials requires a long and costly process of chemical experimentation by trial and error. In recent years, with the rapid development of machine learning computer hardware and software technology, the use of machine learning to accelerate the discovery of new materials has become a hot spot in the field of materials science research\cite{liu2021machine}, and the corresponding newly developed discipline is called as Materials Genome Initiative (MGI), whose core idea is to realize high-speed screening of tens of millions of new materials in chemical space by building high-precision machine learning models based on massive existing materials data.

Our project is based on the idea of MGI, and based on the hybrid organic-inorganic double perovskites (HOIDP) collected from the literature, we use several classical machine learning (ML) algorithms to model the formation energies for chemical stability, energy bandgaps for photovoltaic performances, and Debye temperatures for thermal transport properties, and then use these ML models to screen out new HDOIP candidates materials with excellent chemical stability, photovoltaic performances, and thermal properties from more than 180,000 electrically neutral potential HOIDP chemical spaces\cite{cai2022discovery}. We also use the powerful and physically-inspired CGCNN method to further predict the bandgaps of filtered out HOIDP candidates via the ML models for formation energies and bandgaps, and then use \textit{ab-initio} method to calculate the bandstructures of randomly chosen HOIDP candidates from those filtered out by the CGCNN method, for verification of our CGCNN-prediction. The HOIDP photovoltaic materials screened out in this work will provide a rich research object for the future discovery of high performance environmentally friendly and thermally stable perovskite photovoltaic materials.

\section{Numerical Methods}
\subsection{Machine-learning algorithms}

The machine learning methods are computer sciences based on large amounts of data to construct probabilistic models for prediction and analysis of data, and in the recent decade the ML methods have been widely used in the field of materials science research. The ML algorithms can be classified into classification and regression models based on data characteristics, and in this work we use both regression models and classification models for different purposes. The ML algorithms listed below are summarized based on Ref.\cite{scikit}.

\subsubsection{Methods}
\textbf{Gradient Boosting} The Gradient Boosting (GB) is a class of algorithms in Boosting, which borrows its idea from gradient descent. Its basic principle is to train newly added weak classifiers based on the negative gradient information of the loss function of the current model, and then combines the trained weak classifiers into the existing model in the form of accumulation. The detailed algorithm of GB is listed as below,

\begin{breakablealgorithm}
\caption{Gradient Boosting}
\begin{algorithmic}[1]

\begin{gather}
F_0(x) = \mathop{\arg\min}\limits_{\gamma}\sum^{n}_{i=1}L(y_i, \gamma) \\
L = (y_i - \gamma)^2 
\label{f0}
\end{gather}

\STATE $F_0(x)$ is the target function which predicts a value from $x$, where $x$ is the sample we input.
\STATE $L(y_i, \gamma)$ is the loss function. 
\FOR{m = 1 to M}
\STATE (M is the number of trees we are creating and the small m represents the index of each tree)
\STATE \begin{align}
r_{im} &= -	\left[ \frac{\partial L(y_i, F(x_i))}{\partial F(x_i)} \right]_{F(x)=F_{m-1}(x)} \\
&=-\frac{\partial (y_i-F_{m-1})^2}{\partial F_{m-1}} \notag\\
&=2(y_i-F_{m-1}) \notag 
\end{align}
\ENDFOR
\STATE $F_{m-1}$ is the predicted value from the previous step.
\STATE \textbf{Train regression tree with features $x$ against $r$ and create terminal node residuals $R_{jm}$ for $j = 1,...,J_m$} $j$ represents a terminal node (i.e. leave) in the tree. J is the total numbers of leaves
\STATE \textbf{compute} 

\begin{align}
\gamma_{jm} &= \mathop{\arg\min}\limits_{\gamma} f(x) \sum_{x_i\in R_{jm}}L(y_i, F_{m-1}(x_i)+\gamma) \emph{for} j = 1,...,J_m \\
&= \mathop{\arg\min}\limits_{\gamma} \sum_{x_i\in R_{jm}} (y_i - F_{m-1}(x_i) - \gamma)^2 \notag
\end{align}
\STATE Then, we are finding $\gamma_{jm}$ that makes the derivative of $\sum(*)$ equals zero.
\begin{align}
    &\frac{\partial}{\partial y} \sum_{x_i\in R_{jm}} (y_i - F_{m-1}(x_i)-\gamma)^2 = 0 \\
    &-2 \sum_{x_i\in R_{jm}} (y_i - F_{m-1}(x_i)-\gamma)^2 = 0 \\
    &\gamma = \frac{1}{n_j} \sum_{x_i\in R_{jm}} r_{im}
\end{align}
\STATE \textbf{update the model:}

\begin{align}
F_m(x)=F_{m-1}(x)+v\sum_{j=1}^{J_m}\gamma_{jm}(x\in R_{jm})
\end{align}

$v$ is the learning rate between 0 and 1
\end{algorithmic}
\end{breakablealgorithm}


\textbf{Kernel Ridge} The Kernel Ridge Regression (KRR) is an optimized Linear Regression (LR). It adds \textbf{Kernel} and \textbf{Ridge} on the basis of LR. For normal LR, we train the loss function of MSE, defined as,

\begin{align}
\sum^{n}_{i=1} (y_i - Wx_i)^2
\end{align}

where $W$ is the parameters matrix.

Adding Kernel

\begin{align}
\sum^{n}_{i=1} (y_i - \psi(x_i))^2
\end{align}

where $\psi(x)$ means that, every $x_i$ has its unique parameter $\alpha_j$. Thus the loss function can be expanded to,

\begin{align}
\sum^{n}_{i=1} (y_i - \sum^n_{i=1}\alpha_j\kappa(x_i))^2
\end{align}

Adding Ridge is adding a $L_2$ regularization. The loss function becomes

\begin{align}
\sum^{n}_{i=1} (y_i - \sum^n_{i=1}\alpha_j\kappa(x_i))^2 + \lambda\sum^n_{i=1}\sum^n_{j=1} \alpha_i\alpha_j\kappa(x_j, x_i)
\end{align}

\textbf{Decision Tree} A decision tree (DR) is a map of the possible outcomes of a series of related choices. The detailed algorithm of DR is listed as below, 

\begin{breakablealgorithm}
	\caption{Decision Tree}
	\begin{algorithmic}
	\STATE \textbf{Input:} $D$, a dataset of training records of the form ($X$, $Y$).
	\STATE \textbf{Output:} Root node $R$ of a trained decision tree.
	\STATE Create a root node $R$
	\STATE If a stopping criterion has been reached then label R with the most common value.
	\STATE $Y$ in $D$ and output $R$
	\FOR{$X_i$ in $X$}
	\STATE  Find the test $T_i$ whose partition $D_1, D_2...D_n$ performs best according to the chosen splitting metric.
	\STATE Record this test and the value of the splitting metric
	\ENDFOR
	\STATE Let $T_i$ be the best test according to the splitting metric, let $V$ be the value of the splitting metric, and let $D_1, D_2 .... D_n$ be the partition.
	\IF{$V$\textless threshold}
	\STATE Label $R$ with the most common value of $Y$ in $D$ and output $R$
	\ENDIF
	\STATE Label $R$ with $T_i$ and make a child node $C_i$ of $R$ for each outcome $O_i$ of $T_i$.
	\FOR{each out come $O_i$ of $T_i$}
	\STATE Create a new child note $C_i$ of $R$, and label the edge $O_i$
	\STATE Set $C_i = Train Tree(D_i)$
	\ENDFOR
	\STATE Output $R$
	\end{algorithmic}
\end{breakablealgorithm}

\textbf{Random Forest} The Random Forest (RF) is composed of different decision trees, each with the same nodes, but using different data that leads to different leaves. \\

\begin{breakablealgorithm}
	\caption{Random Forest}
	\begin{algorithmic}
	
	\FOR{b=1 to B}
	\STATE Draw a bootstrap sample $Z^*$ of size $N$ from the training data 
	\STATE Grow a random-forest tree $T_b$ to the bootstrapped data, by recursively repeating the following steps for each terminal node of the tree, until the minimum node size $n_{min}$ is reached.
	\begin{ALC@g}
	 \STATE i. Select m variables at random from the p variables.
	 \STATE ii. Pick the best variable/split-point among the m.
	 \STATE iii. Split the node into two daughter nodes.
	\end{ALC@g}
	\STATE Output the ensemble of trees $\{T_b\}^B_1$.
	\STATE To make a prediction at a new point x:
	\STATE  \textit{Regression:} $\hat{f}^B_{rf}(x)=\frac{1}{B}\sum^{B}_{b=1}T_b(x).$
    \ENDFOR
	\end{algorithmic}
\end{breakablealgorithm}

\textbf{K Nearest Neighbors} The K Nearest Neighbors (KNN) is a supervised learning algorithm that can be used in both regression and classification. The KNN algorithm assumes that, the similar data is always close to each other. The KNN predicts by calculating the distance between the test data and all the training points.

\begin{breakablealgorithm}
    \caption{K Nearest Neighbors}
    \begin{algorithmic}
    \STATE Initialize $K$ to the chosen number of neighbors
    \FOR{each example in the data}
    \STATE Calculate the distance between the query example and the current example from the data.
    \STATE Add the distance and the index of the example to an ordered collection
    \ENDFOR
    \STATE Sort the ordered collection of distances and indices from smallest to largest(in ascending order) by the distances
    \STATE Pick the first $K$ entries from the sorted collection
    \STATE Get the labels of the selected $K$ entries
    \STATE Return the mean of the K labels
    \end{algorithmic}
\end{breakablealgorithm}

\textbf{Support Vector Machine} The Support vector machine(SVM) is a supervised learning algorithm. The SVM algorithm creates the best line or decision boundary that can segregate n-dimentional space. 

\begin{align}
    g(x) = w^Tx+b
\end{align}
Where g(x) divided two class of the dataset. 

We maximize $k$(the sum of two shortest distances between two points that belong to different class divided by $g(x)$), such that
\begin{align}
   & -w^Tx + b \geq k \text{ for } d_i == 1 \text{ $d_i$ is the ith point in the data}\\
   & -w^Tx + b \leq k \text{ for } d_i == -1\\
    &\text{Value of $g(x)$ depends upon $\lVert w \rVert$:}  \notag \\
    &\text{1) Keep $\lVert w \rVert = 1$ and maximize $g(x)$} \notag \\
    &\text{2) $g(x)\geq 1$ and minimize $\lVert w \rVert$} \notag
\end{align}

We can formulate the problem as,
\begin{align}
\phi(w) = \frac{1}{2}w^Tw - minimize \\
\text{Subject to } d_i(w^Tx+b)\geq 1 \text{ }\forall \text{ }i
\end{align}

Integrating the constants in Lagrangian form we get,

\begin{align}
\text{Minimize : }& J(w,b,\alpha) = \frac{1}{2}w^Tw - \sum^N_{i=1}\alpha_i d_i (W^Tx_i + b) + \sum^N_{i=1} \alpha_i \\
&\text{Subject to : } \alpha_i \geq 0 \text{ }\forall \text{ } i
\end{align}

Solve the Lagrangian form we get,

\begin{align}
w_0 = \sum^N_{i=1}\alpha_id_ix_i \text{ and }\sum^N_{i=1} \alpha_id_i = 0
\end{align}

\textbf{Multi-layer Perceptron} The Multi-layer Perceptron (MLP) is an artificial Neural Networks (ANNs). The MLP algorithm can be considered as a directed graph, consisting of multiple nodes layers.

\begin{breakablealgorithm}
    \caption{Multi-layer Perceptron}
    \begin{algorithmic}
    \STATE Select synaptic weights and threshold value for nodes.
    \STATE Calculate the induced signal function signals of the network by moving forward through the network layer by layer.
    \STATE Determine the local gradients of the network.
    \STATE Adjusts the weights according to the local gradients until the error reach the threshold.
    \end{algorithmic}
\end{breakablealgorithm}

\begin{figure}[ht!]
\centering
\includegraphics[width=0.9\linewidth]{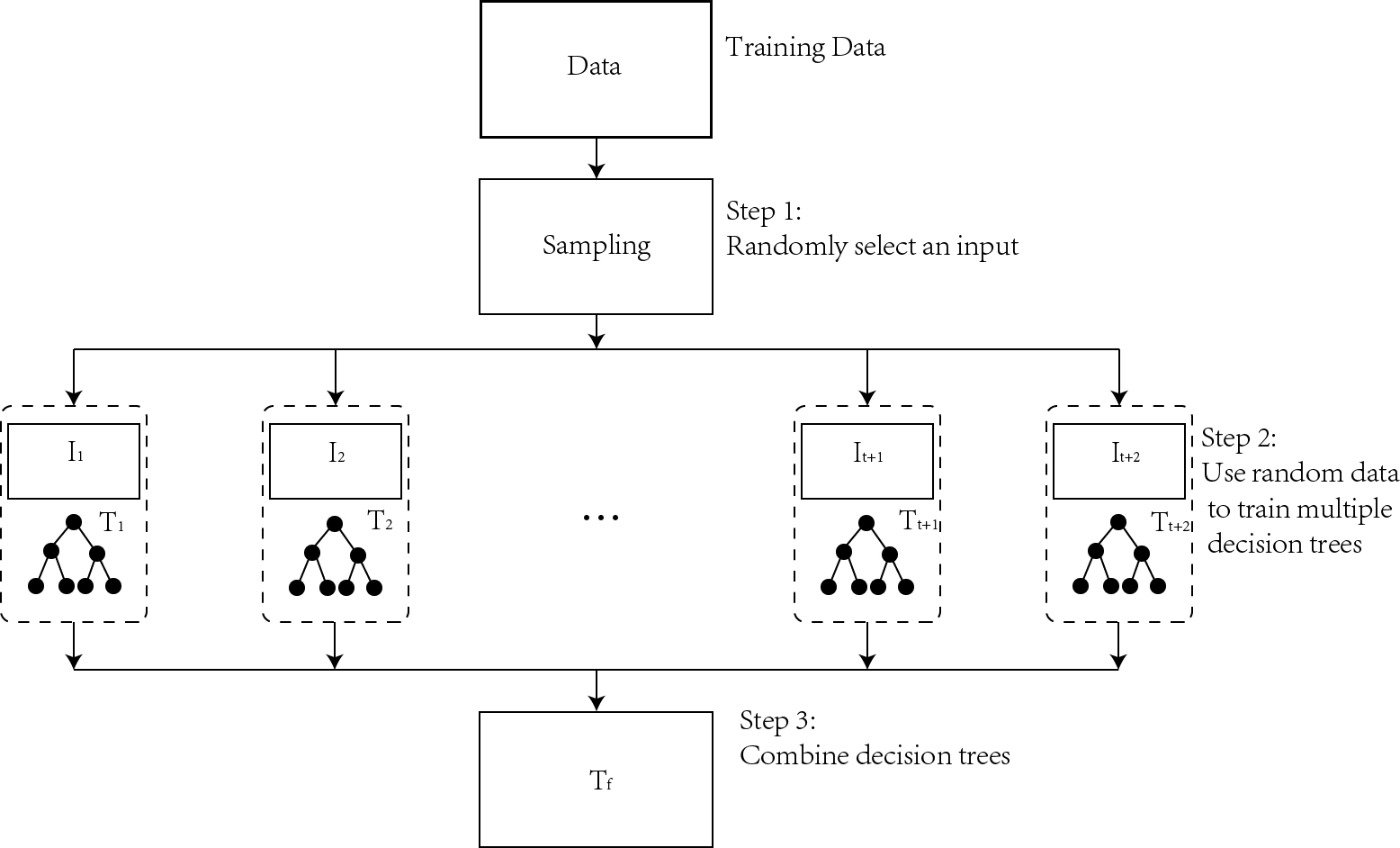}
\caption{The schematic view of ET algorithm.}
\label{fig:et}
\end{figure}

\textbf{Extra Trees Ensemble} As shown in Figure~\ref{fig:et}, the Extra Trees (ET) algorithm is similar to the RF algorithm, but it is different in following aspects, 1) RF uses bootstrap replicas, whereas ET uses the whole sample. By comparison, boostrapping may increase the variance. 2) ET splits the nodes randomly, whereas Random Forest split the node at optimum. Thus, the randomness of ET comes from the node split but not sample selecting.

\begin{figure}[h]
\centering
\includegraphics[width=0.9\linewidth]{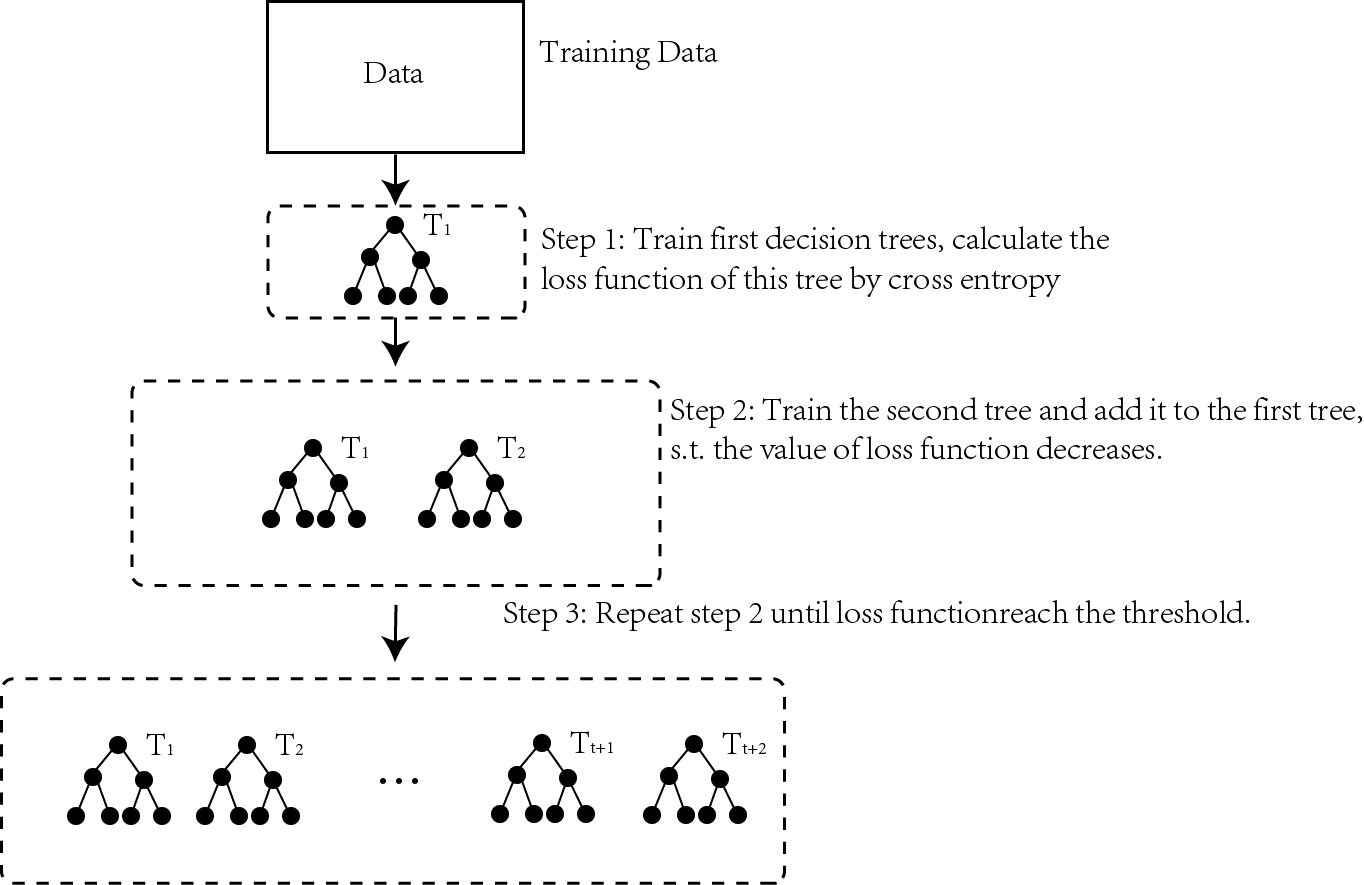}
\caption{The schematic view of eXtreme Gradient Boosting.}
\label{fig:xgboost}
\end{figure}

\textbf{eXtreme Gradient Boosting} As shown in Figure~\ref{fig:xgboost}, the eXtreme Gradient Boosting (XGBoost) algorithm is a boosting algorithm, which produces multiple weak learners to produce a strong learner. XGBoost is different from GB that uses gradient descent to minimize loss function. XGBoost uses the second-order Taylor polynomial to minimize its loss function, i.e. the Newton's method.

\subsubsection{Metrics}

For the regression models, we adopt four metrics, i.e. Coefficient of determination ($R^2$), Mean absolute error (MAE), Mean squared error (MSE), and Cross validation score (CV), which are defined as follows,

\begin{align}
    R^2 =& 1-\frac{\sum^N_{i=1}(y_i-\hat{y_i})^2}{\sum^N_{i=1}(y_i-\overline{y})^2} \\
    MAE& = \frac{1}{N} \sum^N_{i=1}(|y_i-\hat{y_i}|)\\
    MSE& = \frac{1}{N} \sum^N_{i=1}(y_i-\hat{y_i})^2\\
    CV& = \frac{1}{k} \sum^k_{i=1}{MSE}_i\\
\end{align}
where $k$ is number of fold we apply in the cross validation.

For the classification models, we adopt the Accuracy and Precision as the metrics of classification model.
Assuming there are N classes, the object that belong to the $i^{th}$ class is positive to the $i^{th}$ class, and the object that does not belong to the $i^{th}$ class is negative to the $i^{th}$ class. Define $TP_i$ as judging a positive class as positive, $FP_i$ as judging a negative class as positive, $TN_i$ as judging a positive class as negative, and $FP_i$ as judging a negative class as negative. Finally the Accuracy and Precision can be defined as,

\begin{align}
    \text{Precision}_i = \frac{TP_i}{TP_i+FP_i}\\
    \text{Accuracy}_i = \frac{TP_i}{TP_i+TN_i}\\
\label{eq:accuracy}
\end{align}

\subsubsection{Feature-importance Analysis}

Generally, for a better understanding of features to reveal the underlying mechanisms of the targeted property, or reducing dimension of input features to avoid the overfitting problem, it is necessary to calculate and analyze the feature importance. To analyze the features importance in models, in this work we adopt the SHAP (SHapley Additive exPlanation) method to calculate features importance\cite{shap}. The SHAP method calculates the Shapley value of each features, as well as the influence of changing that feature on the output one by one. Therefore, by using SHAP, we not only can obtain the most important features that determine the targeted property, but also reveal the dependence of the targeted property on each changing feature.

\subsection{Crystal Graph Convolution Neural Network (CGCNN) method}
As we know, the physical properties of materials are mainly determined by the attributes of constituent atoms and the crystal structures. However, the above-mentioned classical ML algorithms generally use atomic attributes as the input features to construct the ML models for the purpose to predict some special properties of materials, which in principle lacks of intrinsic precision due to the neglect of information of crystal structures. Therefore, the CGCNN method was proposed based on the recent developed graph neural network (GNN)\cite{tian2018}.

\begin{figure}[ht!]
\centering
\includegraphics[width=0.9\linewidth]{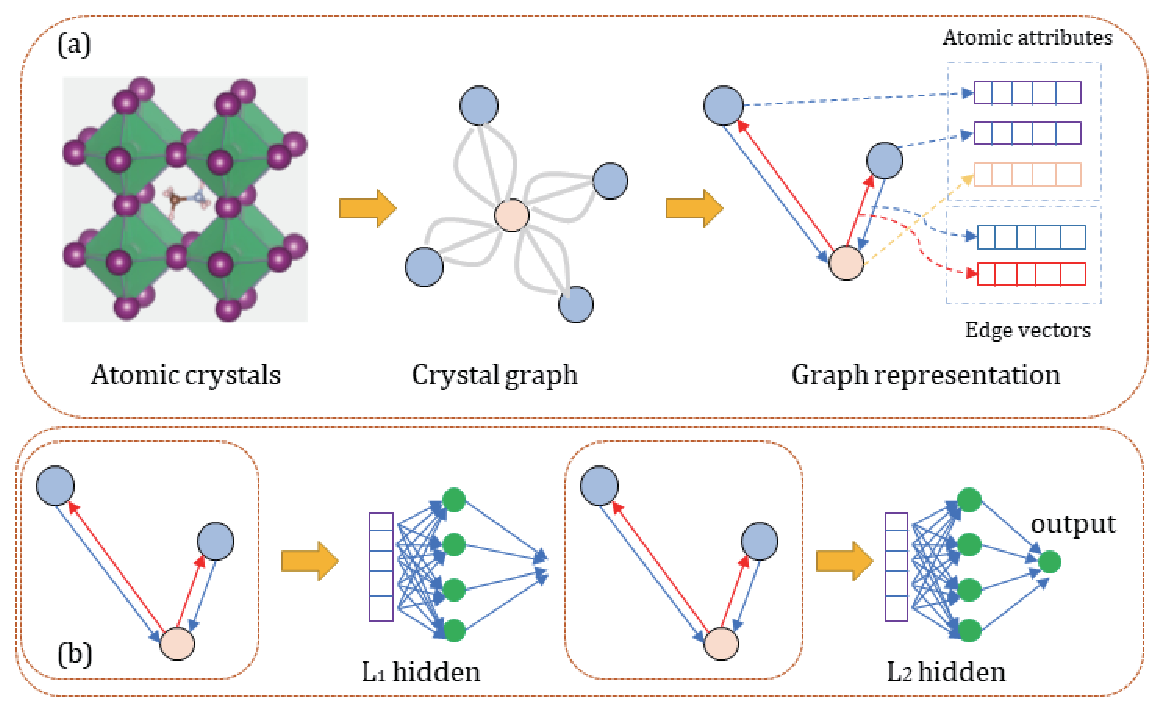}
\caption{The CGCNN algorithm.}
\label{fig:cgcnn}
\end{figure}

The CGCNN method, represents the real crystal structures by crystal graph through encoding the atomic attributes and bonding information between atoms, as shown in Figure~\ref{fig:cgcnn}(a), and then a convolutional neural network is built on top of the crystal graph, as shown in Figure~\ref{fig:cgcnn}(b), for the prediction of targeted property by training the input data.
~Overall the relative advantages of the CGCNN method include, 1) the structure of a Graph Convolution Neural Network is more close to real material structures, so it is more explicable than other Machine Learning algorithms; 2) CGCNN adopts CIF (Crystallographic Information File) as its input, which is a standard representation of materials in computer. CIF is much easier to attain than making CSV file as input; 3) CGCNN adopts Neural Network, i.e. having every advantages of Neural Network over Classic Machine Learning Methods.

\subsection{\textit{ab-initio} calculation methods}

The calculations regarding the electronic bandstructures are implemented using the Vienna Ab Initio simulation package (VASP) based on density functional theory (DFT)\cite{Kresse1996}. The generalized gradient approximation (GGA) proposed by Perdew, Burke, and Ernzerhof (PBE) is utilized for the exchange-correlation energy\cite{Perdew1996}. The calculation is carried out by using projector-augmented-wave (PAW) pseudopotential method with a kinetic energy cutoff of 500 eV. The energy convergence value between two consecutive steps is set to be $10^{-3}\;$eV when optimizing atomic positions and the Hellman–Feynman force convergence threshold is $10^{-1}\;$eV/\AA. The Brillouin zone is sampled by a grid of $2\times2\times2$ k mesh. 

\section{RESULTS and DISCUSSIONS}

By using the classical ML algorithm and CGCNN methods, the formation energies, bandgaps and Debye temperatures of a series of HOIDPs are modelled with high precision, and then used to predict and screen out potential HOIDPs with high performances beneficial for photovoltaic applications. Firstly, we use classical regression algorithms to model the formation energies of HOIDPs, and then use the chosen regression models to screen a built dataset to filter out thermally stable HOIDPs with the formation energies less than zero. Secondly, we build the regression models for bandgaps, as well as the classification models with three categories to screen the thermally stable HOIDPS and filter out those with appropriate bandgaps beneficial for photovoltaic applications. We also use CGCNN method to predict the bandgaps for those formerly classified as good photovoltaic materials and screen out those with bandgaps within the range of 0.5 eV to 2.0 eV, which are subsequently verified by the DFT calculations. Finally, we build the regression models for Debye temperatures to screen thermally stable HOIDPs with appropriate bandgaps, and filter out those with Debye temperatures higher than 500 K.

\subsection{The machine-learning modelling}

\subsubsection{Data preprocessing}

In this work, three properties of HOIDPs are under investigations, i.e. formation energy, bandgap, and Debye temperature. In total, we have 4456 samples of bandgaps and formation energies\cite{HT-DFT-1,HT-DFT-4,HT-DFT-5}, and 426 samples of Debye temperatures\cite{cai2022discovery}. 36 types of atomic attributes are used as the main features, as listed in Table~\ref{tab:features}. After simple mathematical algebra between features, total 95 types of features are generated. For those missing data, we set them to zero. 

The datasets are devided into the training dataset and the testing dataset. In some ML models, such as the DT model, the normalization process is important for their performance. So we also train the normalizer by training dataset, then apply the same normalizer to testing dataset in case of model that can learn the distribution of testing dataset in the training process.

\begin{table}[ht!]
\centering
\caption{Main features based on atomic attributes.}
\begin{tabular}{ cccc  }
 \hline
 Feature & Meaning & Feature & Meaning\\
 \hline
 HOMO & highest occupied molecular orbital of A site & $M_i$ & Atomic weight\\
 LUMO &  lowest unoccupied molecular orbital of A site & $R_i^{CON}$ &Covalent radius\\
 HOMO-LUMO & Difference between HOMO and LUMO & $P^d_i$ & Dipole Polarizability\\
 $H_i^{ev}$ & Evaporation heat & $D_i$ & Thermal conductivity\\
 $R_B^{ion}$ & Average of B-site ionic radius & $R_i^{ion}/R_j^{ion}$ &Ratio of two effective ionic radius\\
 $R_i^{ion}+R_j^{ion}$ & Sum of two effective ionic radius & $R_i^{ion} - R_j^{ion}$ & Difference of two effective ionic radius\\
 $P_A$ & Ionic polarizability of A site. & $E_i^{ea}$ & Electron affinity\\
 $R_i^{ion}$ & Ionic radius & $R_i^{metal}$ & Metallic radius\\
 $E_i^{ip}$ & First ionization energy & $E_i^{pa}$ & Proton affinity\\
 $R_i^{s+p}$ &Sum of $s$ and $p$ pseudo-potential orbital radius & $N_i^{men}$ & Mendeleevs number\\
 $N_i^{atom}$ &Atom number & $N_i^{period}$ & Period number\\
 $R_i^{atom}$ & Atomic radius & $N_i^proton$ & Number of protons\\
 $P^a_A$ & Anisotropic polarizability of A size & $X_i$ & Pauling scale of electronegativity\\
 $N_A$ & The number of atoms in A site & M & Sum of the atom weight\\
 $M_A$ & The weight of A site & $T_i^MP$ & Melting point \\
 $T_f$ & Tolerance factor & $H_i^{fh}$ &Heat of formability\\
 $O_f$ & Octahedral factor & $R_i^{vdw}$ & Van der Waals radius\\
 $V_i$ & Atomic volume & $T_i^{bp}$ & Boiling point\\
 \hline
\end{tabular}
\label{tab:features}
\end{table}

\subsubsection{Training}

In this work, the training process includes the following steps,

\begin{enumerate}

\item Train regression models for formation energies, bandgaps, and Debye temperatures by the above-mentioned GBR, KRR, DTR, RFR, KNN, SVM, MLP, ETR, and XGB models with default hyper-parameter.

\item Choose four best models according to their MAE.

\item Adjust hyper-parameters of each models minutely to reach a lower MAE. To achieve that, we set minus score of ten cross-validation as objective function, then minimize the objective function by changing hyper-parameter.

\item We then use the GBC, KRC, DTC, RFC, KNN, SVM, MLP, ETC, and XGC models to make classification models for bandgaps. We adjust hyper-parameters of four models with higher F1-scores.
\end{enumerate}

\subsubsection{Results and analysis}

The performances of nine models for formation energies of HOIDPs are shown in Table~\ref{tab:formation}, which shows that, the performances of XGB, RFR, ETR, and GBR are the best four models. We choose XGB, RFR, ETR and GBR models to optimize their hyper-parameters, and the results are listed in Table~\ref{table:10}, which shows that, all the four models have high $R^2$ scores, i.e. 99.1\%, 98.0\%, 97.9\% and 99.1\%, high enough for further prediction.

\begin{table}[h!]
\caption{Metrics on Regression Models of Formation Energy}
\label{tab:formation}
\begin{tabular}{ cccccccccc }
 \hline
 & GBR&KRR&DTR&RFR&KNN&SVM&MLP&ETR&XGB\\
 \hline
 R2&0.974&0.752&0.963&0.981&0.773&0.541&0.841&0.979&0.986\\
 MAE &0.025&0.062&0.0298&0.019& 0.068&0.050&0.050& 0.018&0.015\\
 MSE &0.0013&0.0119&0.0018&0.0009&0.0109&0.0221& 0.0061&0.0010&0.0006\\
 \hline
\end{tabular}
\end{table}

\begin{table}[h!]
\centering
\caption{Metrics of Adjusted Models}
\label{table:10}
\begin{tabular}{ ccccc  }
 \hline
 & XGB& RFR & ETR & GBR\\
 \hline
 cross val score& 0.987& 0.975&0.984&0.983\\
 R2&0.991&0.980& 0.979&0.991\\
 MAE &0.013&0.019&0.017& 0.013\\
 MSE &0.00045&0.00098&0.00099&0.00045\\
 \hline
\end{tabular}
\end{table}

\begin{figure}[ht!]
\centering
\includegraphics[width=0.9\linewidth]{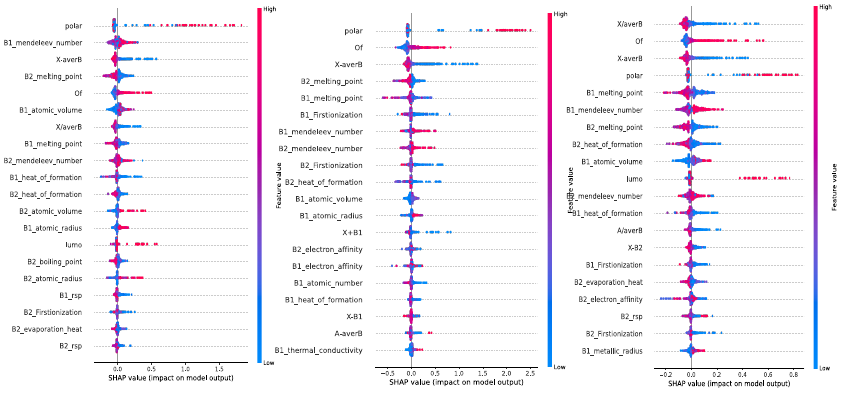}
\caption{The SHAP analysis for bandgaps based on (a) ETR, (b) XGB, and (c)RFR models.}
\label{fig:bandgap-shap} 
\end{figure}

After optimizing hyper-parameters, we calculate and analyze features of importance for ETR, XGB, RFR and GBR by calculating their SHAP values, and the results are shown in Figure~\ref{fig:bandgap-shap}(a-d), respectively. Figure~\ref{fig:bandgap-shap} reveals that, the most influential features for formation energies of HOIDPs are electron affinity of element in X site, Mendeleev number of element in $B_1$ site, first ionization energy of element in $B_1$ site, and heat of formation of element in X site.

\begin{figure}[ht!] 
\centering
\includegraphics[width=0.9\linewidth]{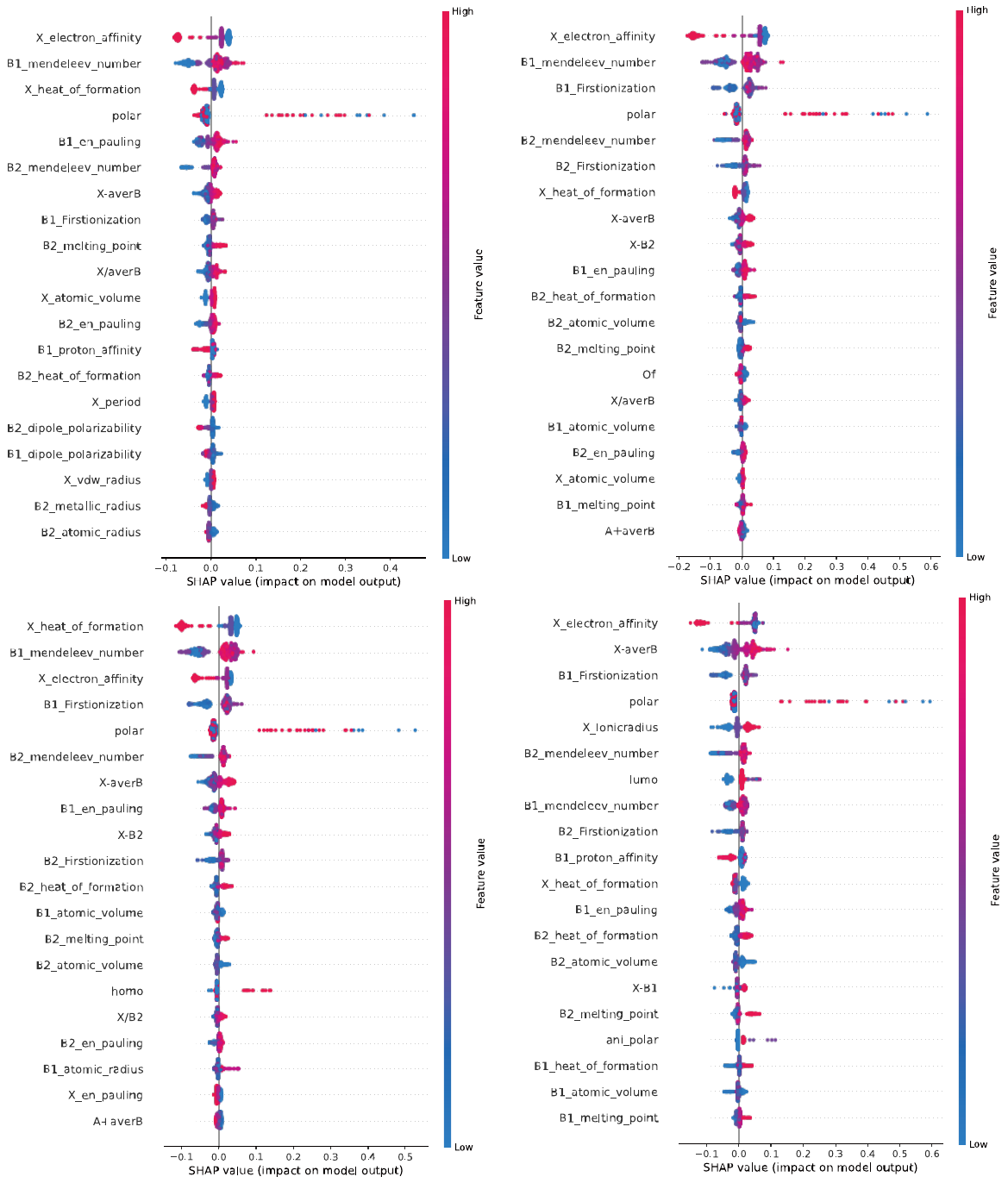}
\caption{The SHAP analysis for formation energies based on (a) ETR, (b) XGB, (c) RFR and (d) GBR models.}
\label{fig:formation-shap} 
\end{figure}

In addition, there are some features that show strongly positive or negative relationship with formation energies. The electron affinity of element in X site and heat of formation of element in X site have a negative relationship with formation energies, while the first ionization energy of element in $B_1$ site and Mendeleev number of element in $B_1$ site have a positive relationship with formation energies. This finding obtained by purely ML modelling is insightful, since, as we know, the stabilities of perovskites can be roughly determined by the appropriate values of Goldschmidt tolerance factors ($T_f$) and octahedral factors ($O_f$), which are defined by\cite{burger2018tolerance},

\begin{equation}
T_f=\frac{r_A+r_X}{\sqrt{2}(r_B+r_X)}
\end{equation}
\begin{equation}
O_f=\frac{r_B}{r_X}
\end{equation}

where $r_{A/B/X}$ denotes the radius of A/B/X-site atoms. As previously reported, stable perovskites generally possess $T_f$ values of $0.81\sim 1.11$, and $O_f$ values of $0.44\sim 0.90$\cite{burger2018tolerance}, respectively. Such an agreement also indicates the precision of our ML modelling for formation energies.

The performances of the nine models for bandgaps are shown in Table~\ref{tab:bandgap}, which shows that, the performances of ETR, XGB, RFR, and DTR models are the best four models. It is noteworthy that MAE of ETR is almost equal to 0.1 eV. We choose the ETR, XGB, RFR and MLP models as the models of which we are going to optimize the hyper-parameters, because MLP is different in principle from ETR, XGB, RFR and DTR that are based on tree model.

\begin{table}[ht!]
\centering
\caption{Metrics on Regression Models of Bandgaps.}
\label{tab:bandgap}
\begin{tabular}{ cccccccccc  }
 \hline
 & GBR&KRR&DTR&RFR&KNN&SVM&MLP&ETR&XGB\\
 \hline 
 $R^2$&0.732&0.487&0.892&0.816&0.578&0.181&0.841&0.940&0.919\\
 MAE &0.308&0.496&0.166&0.156&0.348&0.353&0.269&0.107&0.147\\
 MSE &0.274&0.526&0.111&0.189&0.433&0.840&0.163&0.061&0.083\\
 \hline
\end{tabular}
\end{table}

In searching the best hyper-parameters, we do not use MAE as the metrics. Instead, we adopt score of cross validation for more robust models. The metrics of the best four models is lised in Table~\ref{table:3}, which shows that, although ETR and RFR have a higher MAE after optimizing hyper-parameters, MSE of all models drops obviously. In other words, the spread of data are more stable after adjusting hyper-parameters. The $R^2$ values for the chosen ETR, XGB, RFR and MLP models reach 93.5\%, 93.1\%, 89.8\% and 85.9\%, respectively.

\begin{table}[h!]
\centering
\caption{Metrics of Adjusted Models for Bandgaps.}
\label{table:3}
\begin{tabular}{ ccccc  }
 \hline
 & ETR& XGB & RFR& MLP\\
 \hline
 cross val score& 0.938& 0.903&0.880&0.888\\
 $R^2$&0.935&0.931&0.898&0.859\\
 MAE &0.110&0.127&0.168&0.249\\
 MSE &0.067&0.071&0.105&0.144\\
 \hline
\end{tabular}
\end{table}

We also calculate and analyze the importance of each feature by the SHAP value based on the ETR, XGB and RFR models, as shown in Figure~ \ref{fig:bandgap-shap}(a-c), respectively.
Because of the particular structure of MLP, we can not calculate features importance in a MLP model. For these figures, there are blue dots and red dots, the more red it is, the larger contribution to the targeted property of bandgaps this feature is. The more blue it is, the smaller contribution this feature is. Figure~ \ref{fig:bandgap-shap} reveals that, the most important features for bandgaps of HOIDPs are molecular polarizability, Octahedral factor, the mendeleev number of first element in B site, and difference between volume of atom in X site and average volume of atoms in B site. Since all important features are related to X site and B site, which suggests that bandgaps of HOIDPs are related to elemental properties in X site and B site. This finding is in good consistence with the previous studies on the formation of bandgaps in hybrid perovskites, which reported that, the bandgaps of some conventional hybrid perovskites such as CH$_3$NH$_3$PbX$_3$ (X=F/Cl/Br/I) are formed by the valence band attributed from $X-p$ and $B/B^\prime-s$ atomic orbitals and the conduction band attributed from $X-p$ and $B/B^\prime-p$ atomic orbitals. 

Apart from the regression models, we also build the classification models for bandgaps to identify the materials with bandgaps appropriate for photovoltaic applications. It is well known that semiconductors with bandgaps around 1.34 eV can absorb a large part of sunlight energy, which is beneficial for photovoltaic applications. Since the PBE bandgaps in our dataset are always smaller than real ones, therefore, in this work, we use bandgaps within the range of 0.5 eV to 2.0 eV as the appropriate bandgaps. We classify HOIDPs into three categories regarding their bandgaps, i.e. $\textless$ 0.5 eV, 0.5 eV $\leq$ bandgap \textless 2.0 eV, and $\geq$ 2.0 eV.

\begin{table}[ht!]
\centering
\caption{Metrics on 3-Class Classification Models of Bandgap.}
\label{tab:classification}
\begin{tabular}{ cccccccc  }
 \hline
 & GBC&RFC&DTC&KNN&MLP&ETC&XGB\\
 \hline
 Accuracy&0.886&0.936&0.917&0.815&0.907&0.926&0.927\\
 Precision&0.781&0.881&0.809&0.589&0.806&0.849&0.854\\
 \hline
\end{tabular}
\end{table}

\begin{figure}[ht!]
\centering
\includegraphics[width=1\linewidth]{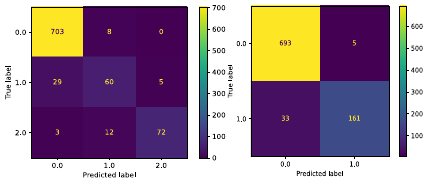}
\caption{Calculated confusion matrix by XGB for (a) 3-categories and (b) 2-categories.}
\label{fig:confusion} 
\end{figure}

The performances of classification models for bandgaps are listed in Table~\ref{tab:classification}, which shows that, RFC, XGBoost, and ETC are still high-precision models for classifying HOIDPs with appropriate bandgaps.
We choose the XGB model to calculate the confusion matrix for the trained dataset, and the results are shown in Figure~\ref{fig:confusion}, which indicates that, the classification models still have a lowest accuracy on classifying label 1.0 (0.5 eV $\leq$ bandgap \textless 2.0 eV). However, classifying label 1.0 correctly is the most important for identifying excellent photovoltaic materials. Considering that there are a lot of metallic materials that have no bandgap, we chose to classify perovskite materials into two categories, i.e. bandgap $\leq$ 0 and bandgap $\textgreater$ 0. The metrics is listed in Table~\ref{table:5}. The Confusion Matrix of XGBoost for both three- and two-category schemes is shown in Figure~\ref{fig:confusion}. With only 5 mistakes when classifier outputs 1.0, we can filter out perovskite candidates with bandgaps larger than 0 eV for further prediction via the CGCNN method.

\begin{table}[htbp]
\centering
\caption{Metrics on 2-class Classification Models of Bandgap}
\label{table:5}
\begin{tabular}{ cccccccc }
 \hline
 & GBC&RFC&DTC&KNN&MLP&ETC&XGB\\
 \hline
 Accuracy&0.889&0.957&0.930&0.808&0.929&0.955&0.954\\
 Precision&0.870&0.962&0.905&0.719&0.921&0.950&0.951\\
 \hline
\end{tabular}
\end{table}


Finally, we build the regression models for Debye temperatures of HOIDPs, and the performances of nine models are listed in Table~\ref{tab:debye}, which shows that, MLP, GBR, XGB and ETR models are the best four models. It is also noteworthy that the MLP model has the best metrics. 

\begin{table}[h!]
\centering
\caption{Metrics on Regression Models of Debye Temperature}
\label{tab:debye}
\begin{tabular}{ cccccccccc  }
 \hline
 & GBR&KRR&DTR&RFR&KNN&SVM&MLP&ETR&XGB\\
 \hline
 R2&0.990&0.982&0.978&0.962&0.954&0.618&0.993&0.987&0.987\\
 MAE &14.3&21.1&21.5&28.1&32.7&72.9&13.0&16.9&15.9\\
 MSE &449.7&809.7&991.4&1670.4&2007.4&16823&298.7&590.6&553.4\\
 \hline
\end{tabular}
\end{table}

\begin{table}[h!]
\centering
\caption{Metrics of Adjusted Models}
\label{tab:debye2}
\begin{tabular}{ccccc  }
 \hline
 & MLP& GBR & XGB & ETR\\
 \hline
 cross val score& 0.992& 0.989&0.984&0.971\\
 R2&0.993&0.993& 0.991&0.990\\
 MAE &13.0&13.3&13.2&15.0\\
 MSE &298.6&326.9&386.1&397.2\\
 \hline
\end{tabular}
\end{table}

Similar to the above-mentioned method, we also optimize the hyper-parameters for the chosen MLP, GBR, XGB and ETR models, and the results are listed in Table~\ref{tab:debye2}, which shows that, $R_2$ values for them are 99.3\%, 99.3\%, 99.1\% and 99.0\%, respectively. For simplificaition, we decide to use MLP to screen materials.

\begin{figure}[ht!]
\centering
\includegraphics[width=1\linewidth]{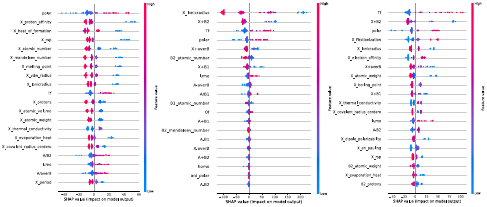}
\caption{The SHAP analysis for Debye temperatures based on (a) ETR, (b) XGB, and (c) GBR models.}
\label{fig:debye-shap} 
\end{figure}

After adjusting hyper-parameters, we calculate and analyzes features of importances for ETR, XGB, and GBR models by SHAP, and the result are shown in Figure~\ref{fig:debye-shap}, which reveals that, the most influential features are radius of ion in X site, tolerance factor, molecular polarizability, and sum of volume of element in X site and $B_2$ site. It is noteworthy that there are many features showing positive relationship or negative relationship with Debye temperatures. For instance, molecular polarizability and Tolerance factor have a positive relationship with Debye temperatures. The sum of volume of element in X site and $B_2$ site and radius of ion in X site have a negative relationship with Debye temperatures.

\subsubsection{Screening HOIDPs based on formation energies, bandgaps and Debye temperatures}

\begin{figure}[ht!]
\centering
\includegraphics[width=0.95\linewidth]{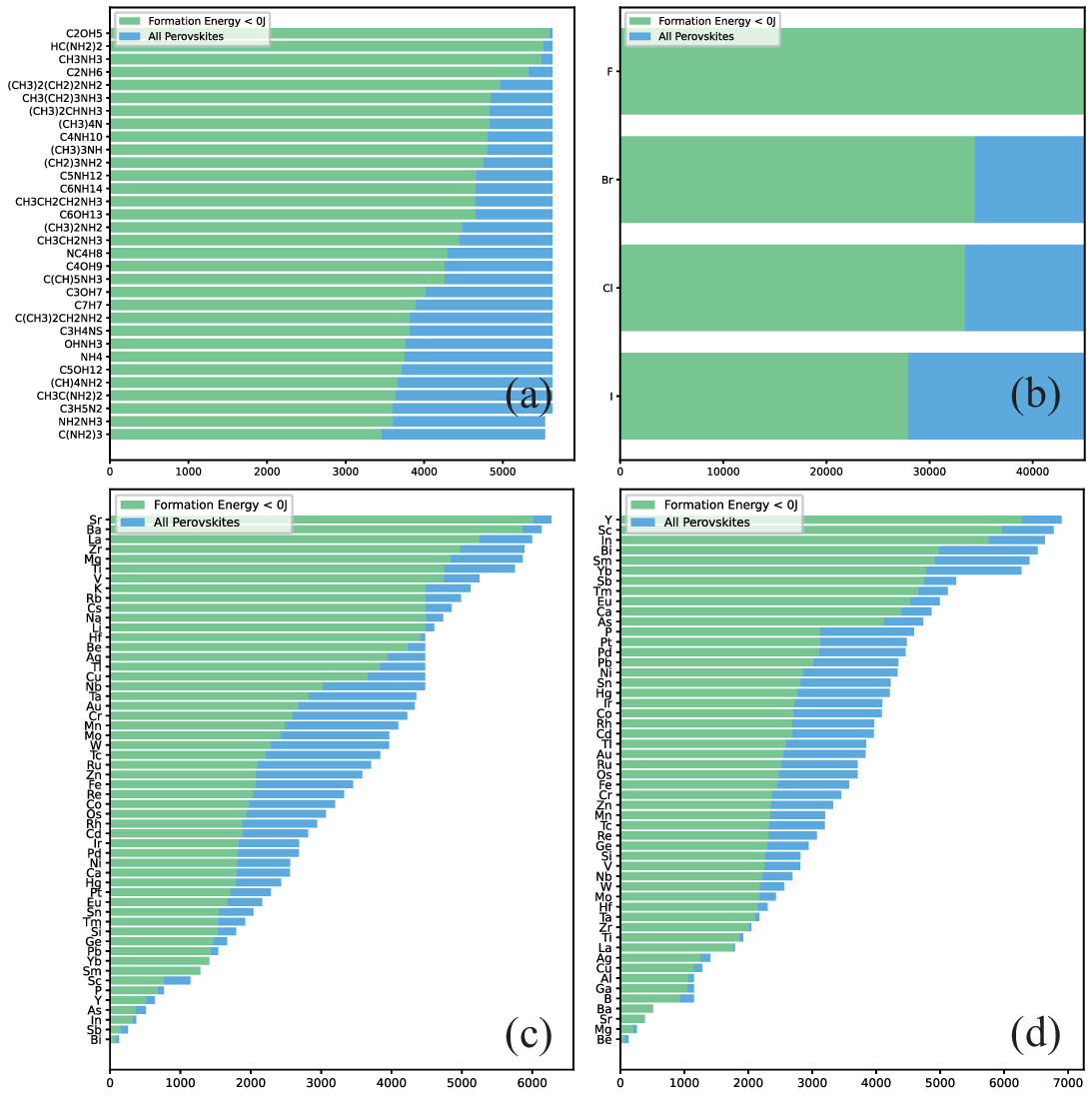}
\caption{The distribution regarding (a) A-, (b) X, (c) $B_1-$ and (d) $B_2-$site elements of the screened out HOIDP candidates with formation energies less than 0 eV/atom. (unit of x-axis is the quantity of compounds)}
\label{fig:screen-formation} 
\end{figure}

We use the GBR model to screen 180,037 HOIDP candidates\cite{HT-DFT-1,HT-DFT-4,HT-DFT-5}, and 140,775 of them are then filtered out with the formation energies less than 0 eV/atom. The distribution regarding A-, X, $B_1-$ and $B_2-$-site elements are shown in Figure~\ref{fig:screen-formation}(a-d), respectively. 

According to Figure~\ref{fig:screen-formation}(a), (CH$_3$)$_2$NH$_2$, (CH$_3$)$_2$(CH$_2$)$_2$NH$_2$, CH$_3$(CH$_2$)$_3$NH$_3$, C$_2$OH$_5$, C(CH)$_5$NH$_3$, CH$_3$CH$_2$NH$_3$, C$_2$NH$_6$, HC(NH$_2$)$_2$, C$_7$H$_7$, and CH$_3$CH$_2$CH$_2$NH$_3$ are identified as the most frequently chosen organic groups, and ten most frequently chosen $B_1-$ and $B_2-$site elements are Be, Tl, Rb, Pd, Cs, Pt, Pd, Os, Ru, and B respectively, as shown in Figure~\ref{fig:screen-formation}(c-d). F-based HOIDPs seem much more stable compared to  Cl/Br/I-based ones, as shown in Figure~\ref{fig:screen-formation}(b).


\begin{figure}[ht!]
\centering
\includegraphics[width=0.95\linewidth]{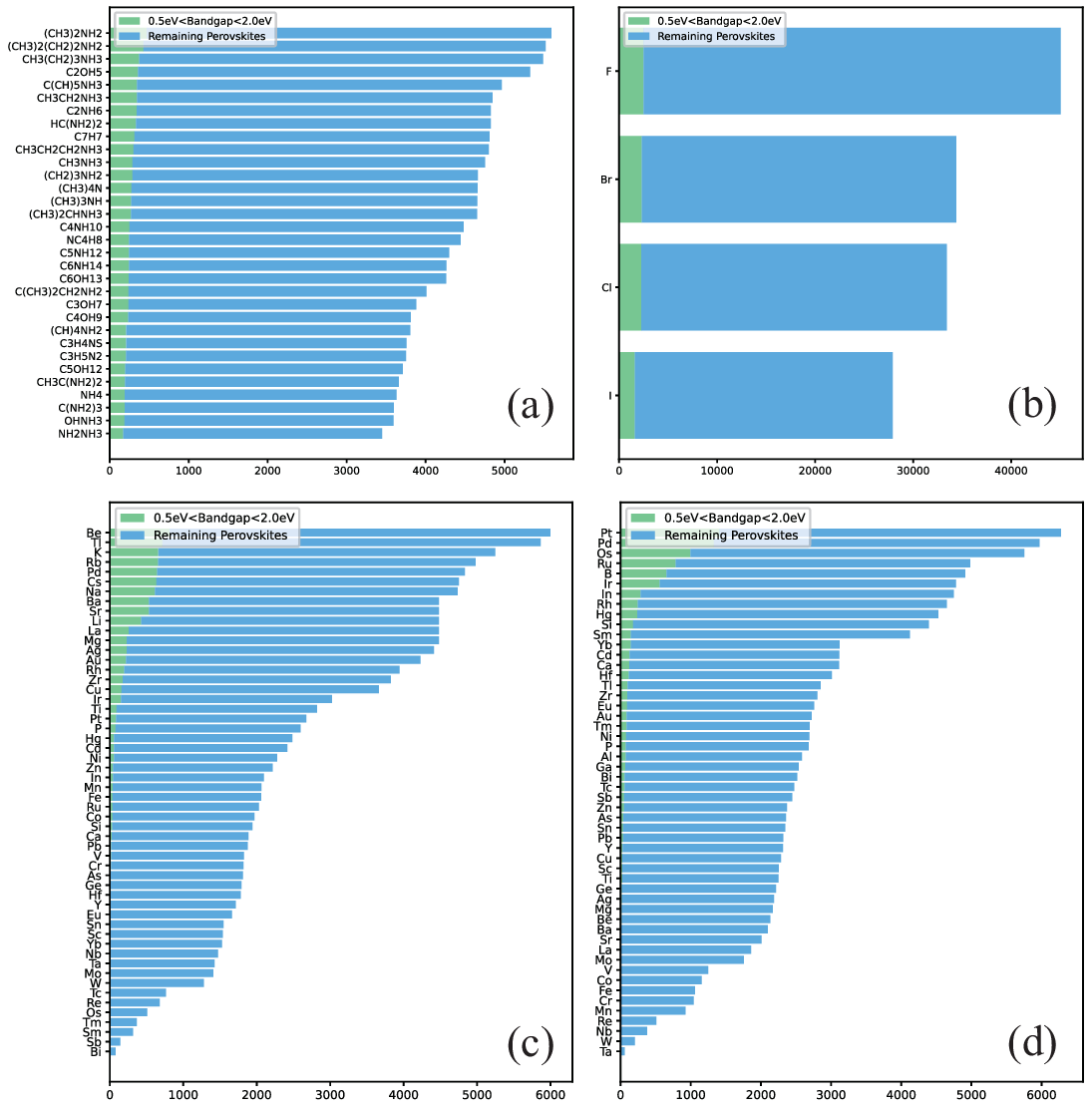}
\caption{The distribution regarding (a) A-, (b) X, (c) $B_1-$ and (d) $B_2-$site elements of the screened out HOIDP candidates with bandgaps in the range from 0.5 eV to 2.0 eV. (unit of x-axis is the quantity of compounds)}
\label{fig:screen-bandgap} 
\end{figure}

By using the built RFC classification model to screen 140,775 HOIDP candidates with formation energies less than 0 eV/atom, 8,693 of them with bandgaps in the range from 0.5 eV to 2.0 eV are filtered out. Only nearly 6\% of HOIDPs candidates survive. The distribution regarding A-, X, $B_1-$ and $B_2-$site elements are shown in Figure~\ref{fig:screen-bandgap}(a-d), respectively. According to Figure~\ref{fig:screen-bandgap}(a), (CH$_3$)$_2$NH$_2$, (CH$_3$)$_2$(CH$_2$)$_2$NH$_2$, CH$_3$(CH$_2$)$_3$NH$_3$, C$_2$OH$_5$, C(CH)$_5$NH$_3$, CH$_3$CH$_2$NH$_3$, C$_2$NH$_6$, HC(NH$_2$)$_2$, C$_7$H$_7$ and CH$_3$CH$_2$CH$_2$NH$_3$ are identified as the most frequently chosen organic groups, and ten most frequently chosen $B_1-$ and $B_2-$site elements are La, Mg, Ag, Au, Rh, Zr, Cu, Ir, Ti, Pt and Sm, Tl, Zr, Eu, Au, Tm, Ni, P, Al, Ga, respectively, as shown in Figure~\ref{fig:screen-bandgap}(c-d). Similarly, F-based HOIDPs seem much more stable compared to  Cl/Br/I-based ones, as shown in Figure~\ref{fig:screen-bandgap}(b).  


\begin{figure}[ht!]
\centering
\includegraphics[width=0.9\linewidth]{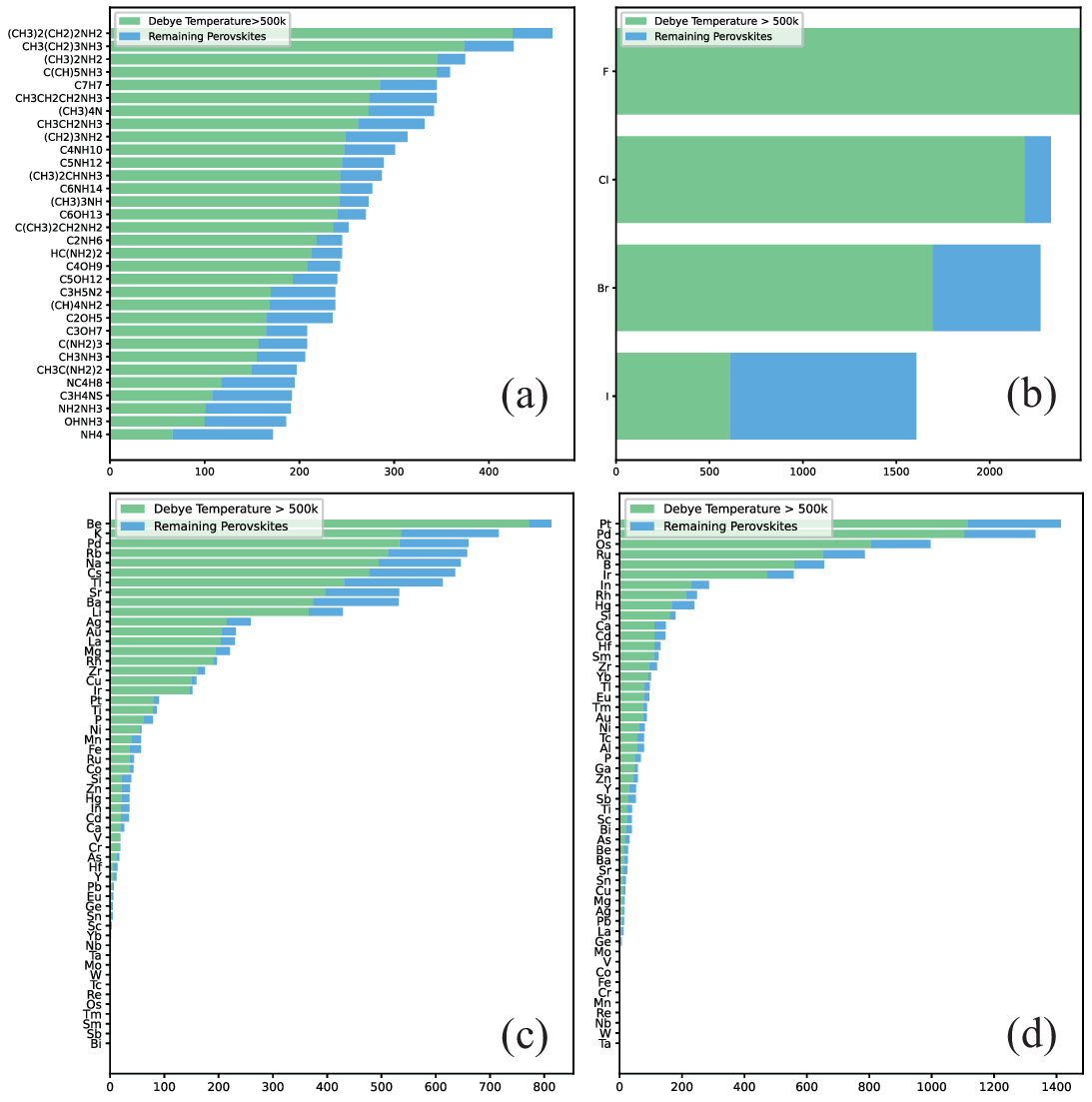}
\caption{The distribution regarding (a) A-, (b) X, (c) $B_1-$ and (d) $B_2-$site elements of the screened out HOIDP candidates with Debye temperatures larger than 500 K. (unit of x-axis is the quantity of compounds)}
\label{fig:screen-debye} 
\end{figure}
Finally, by using the built MLP regression model to screen 8,693 HOIDP candidates, 6,983 of them with Debye temperatures larger than 500 K are filtered out. The distribution regarding A-, X, $B_1-$ and $B_2-$site elements are shown in Figure~\ref{fig:screen-debye}(a-d), respectively. According to Figure~\ref{fig:screen-debye}(a), (CH$_3$)$_2$(CH$_2$)$_2$NH$_2$, CH$_3$(CH$_2$)$_3$NH$_3$, (CH$_3$)$_2$NH$_2$, C(CH)$_5$NH$_3$, C$_7$H$_7$, CH$_3$CH$_2$CH$_2$NH$_3$, (CH$_3$)$_4$N, CH$_3$CH$_2$NH$_3$, (CH$_2$)$_3$NH$_2$, and C$_4$NH$_10$ are identified as the most frequently chosen organic groups, and five most frequently chosen $B_1-$ and $B_2-$site elements are Be, K, Pd, Rb, Na, Pt, Pd, Os, Ru, and B respectively, as shown in Figure~\ref{fig:screen-debye}(c-d). Similarly, F-based HOIDPs seem much more stable compared to  Cl/Br/I-based ones, as shown in Figure~\ref{fig:screen-debye}(b). 


\subsection{The CGCNN Modelling}

After the classical ML modelling, we continue to conduct the CGCNN modelling. 

\subsubsection{Data preprocessing}

Inputs of CGCNN are CIF files. Generally, CIF files contains structure of materials. The first step of preprocessing is to convert structure of a material into vectors. CGCNN create a 93-dimension vector to make one-hot encoding for each atom according to Table~\ref{table:11}.

\begin{table}[ht!]
\centering
\caption{Encoding for Atoms.}
\label{table:11}
\begin{tabular}{ cccc  }
 \hline
 Property & Unit & Range & Classes\\
 \hline
 Group & & 1,2,...18& 18\\
 Period & & 1,2,...9& 9 \\
 Electronegativity && 0.5-4.0 & 10\\
 Covalent Radius & $pm$& 25-250 & 10\\
 Valence Electrons & & 1,2,...,12 & 12\\
 First Ionization Energy &eV& 1.3-3.3&10\\
 Electronic Affinity & eV & -3-3.7&10\\
 Orbital & & s,p,d,f & 4\\
 Atomic Volume & $cm^3/mol$ & 1.5-4.3& 10\\
 \hline
\end{tabular}
\end{table}

where the first column in Table~\ref{table:11} is the chosen atomic attributes. The core of GNN algorithms is not only the nodes, but also the edges. So we also encode chemical bonds between atoms according to Table~\ref{table:12}.

\begin{table}[ht!]
\centering
\caption{Encoding for Bonds.}
\label{table:12}
\begin{tabular}{ cccc  }
 \hline
 Property & Unit & Range & Classes\\
 \hline
 Atomic Distance & $\mathring{A}$ & 0.7-5.2 & 10\\
 \hline
\end{tabular}
\end{table}

\subsubsection{Training}

We use a subset of the above-mentioned 4,456 HOIDPs to generate 3,489 HOIDPs with CIF Files for training. For these 3489 samples, we randomly select 10\% of them as validation set, 10\% of them as test set, and the remaining 80\% as train set. The default hyper-parameters for CGCNN in our case are listed in Table~\ref{tab:CGCNN-parameters}.

\begin{table}[ht!]
\centering
\caption{Grid Optimization of CGCNN for bandgaps.}
\label{tab:CGCNN-parameters}
\begin{tabular}{ccccccccc}
\hline
&Learning Rate&Batch Size& Atom-fea-len&Hidden-fea-len&n-conv&n-hidden&epochs&MAE\\ 
\hline
Default& 0.01& 64 & 64 & 128 &3&1& 30 & 0.303\\
 1&	0.005&	64	&32&	16&	3&	1&	200&	0.193\\
2&	0.01&	64	&32&	16&	3&	1&	200	&0.178\\
3&	0.015&	64&	32	&   16&	3&	1&	200	&0.176\\
4&	0.02&	64&	32&	    16&	3&	1&	200	&0.178\\
5&	0.01&	128&32&	    16&	3&	1&	200	&0.186\\
6&	0.015&	128	&32&	16&	3&	1&	200	&0.192\\
7&	0.02&	128	&32&	16&	3&	1&	200	&0.193\\
8&	0.015&	64&	64&	    16&	3&	1&	200	&0.167\\
9&	0.015&	64&	128&	16&	3&	1&	200	&0.164\\
10	&0.015&	64&	192	&   16&	3&	1&	200	&0.169\\
11	&0.015&	64&	128	&   32&	3&	1&	200	&0.168\\
12&	0.015&	64&	128&	64&	3&	1&	200	&0.163\\
13&	0.015&	64&	128&	128&3&	1&	200	&0.165\\
14&	0.015&	64&	128&	64	&3&	1&	200	&0.163\\
15&	0.015&	64&	128&	64	&4&	1&	200	&0.157\\
16&	0.015&	64&	128&	64	&5&	1&	200	&0.166\\
17&	0.015&	64&	128&	64	&4&	2&	200	&0.162\\
18&	0.015&	64&	128&	64	&4&	3&	200	&0.154\\
19&	0.015&	64&	128&	64	&4&	4&	200	&0.162\\
 \hline
\end{tabular}
\end{table}

We then change hyper-parameter several times, and the results are listed in Table~\ref{tab:CGCNN-parameters}. According to Table~\ref{tab:CGCNN-parameters}, the best hyper-parameters are the 18 row. Although $MAE_{CGCNN}(0.154) \textgreater MAE_{ETR}(0.110)$, CGCNN uses less train set and less features in each train set. This proves that CGCNN is potential to be a better way to predict bandgaps of materials.  

\begin{table}[ht!]
\centering
\caption{Comparison between prediction and experimental results on conventional perovskites.}
\label{tab:CGCNN-performances}
\begin{tabular}{ccc}
\hline
& Experiments (eV)\cite{post-encapsulation} & CGCNN Predict (eV)\\
\hline
FAPbI$_3$& 1.48 & 1.63\\
CH3NH3PbBr$_3$ &2.2&2.03\\
HPbI$_3$ & 1.7 &1.85\\
Cs$_2$AgBiBr$_6$ & 1.98 & 2.23\\
\hline
\end{tabular}
\end{table}

We also use our CGCNN model to predict the bandgaps of some conventional photovoltaic perovskites that are not in our training dataset, and the result are listed in the Table~\ref{tab:CGCNN-performances}, which confirms the precision of our CGCNN model. 

\subsubsection{Screening Materials by CGCNN}

Here, as mentioned above, the CGCNN method may predict properties of materials more precisely than the classical ML algorithms, especially for bandgaps of materials. Since the bandgaps of materials are resulted from many-body interactions among electrons and ions\cite{Good-1}, which is generally very complex and beyond the ability of classical ML algorithms, therefore, we also use the CGCNN method to further predict the bandgaps of those filtered out by the classification models as mentioned above. Then we choose some of them to be verified by the DFT calculations.

\begin{figure}[ht!]
\centering
\includegraphics[width=0.8\linewidth]{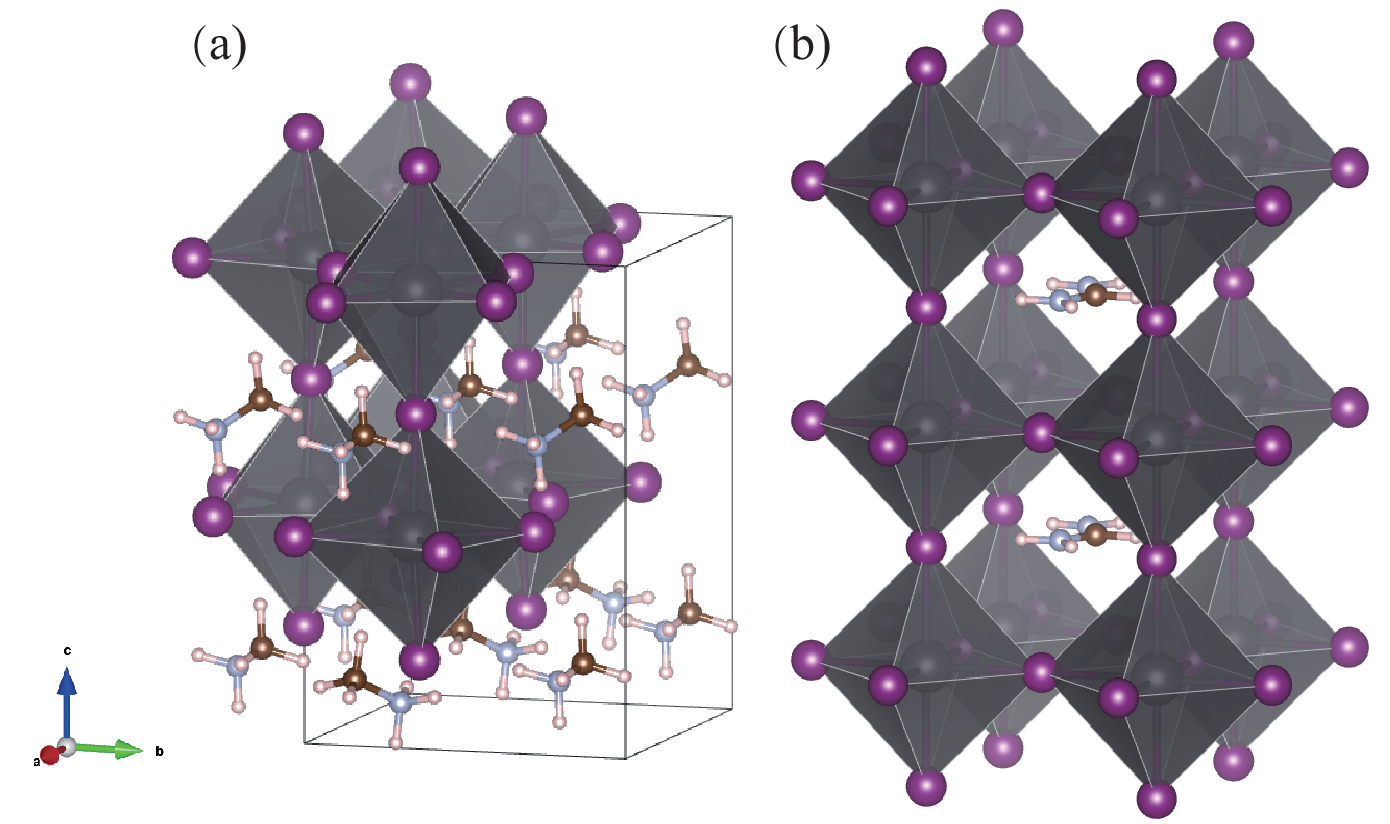}
\caption{Crystal structures of (a) CH$_3$NH$_3$PbI$_3$ with tetragonal phase, (b) HC(NH$_2$)$_2$PbI$_3$ with cubic phase.}
\label{fig:crystals} 
\end{figure}

At room temperatures, hybrid perovskites generally have two phases, i.e. tetragonal and cubic phases\cite{Baikie2013}. The crystal structures of two conventional hybrid perovskites with excellent photovoltaic performances, i.e. CH$_3$NH$_3$PbI$_3$ (MAPbI$_3$) and HC(NH$_2$)$_2$PbI$_3$ (FAPbI$_3$), are shown in Figure~\ref{fig:crystals}(a,b), respectively. Based on the tetragonal and cubic structures as shown in Figure~\ref{fig:crystals}(a,b), a series of crystal structures of potential HOIDP candidates can be subsequently generated for further DFT verification. Herein for simplification and reducing the intense computations, we only consider the HOIDP candidates with tetragonal phase generated based on Figure~\ref{fig:crystals}(a) and replacing the CH$_3$NH$_3$ organic group by 
only one organic groups, i.e. (CH$_3$)$_2$NH$_2$, which are mostly chosen one, as shown in Figure~\ref{fig:screen-bandgap}(a).

\subsection{High-throughput filtering based on bandgaps}

The number of (CH$_3$)$_2$NH$_2$-based HOIDP candidates with tetragonal phase is 365, and 288 of them possess the bandgaps in the range from 0.5 eV to 2.0 eV according to the CGCNN prediction, which is also the range we use in the classification-modelling process to filter out good HOIDP candidates. To reduce the computational intensity, we randomly choose twenty of 288 types of (CH$_3$)$_2$NH$_2$-based HOIDP candidates filtered out based on the CGCNN method, and use the PBE method to calculate their bandstructures, and find that, six of them possess finite bandgaps. 

\begin{figure}[ht!]
\centering
\includegraphics[width=1\linewidth]{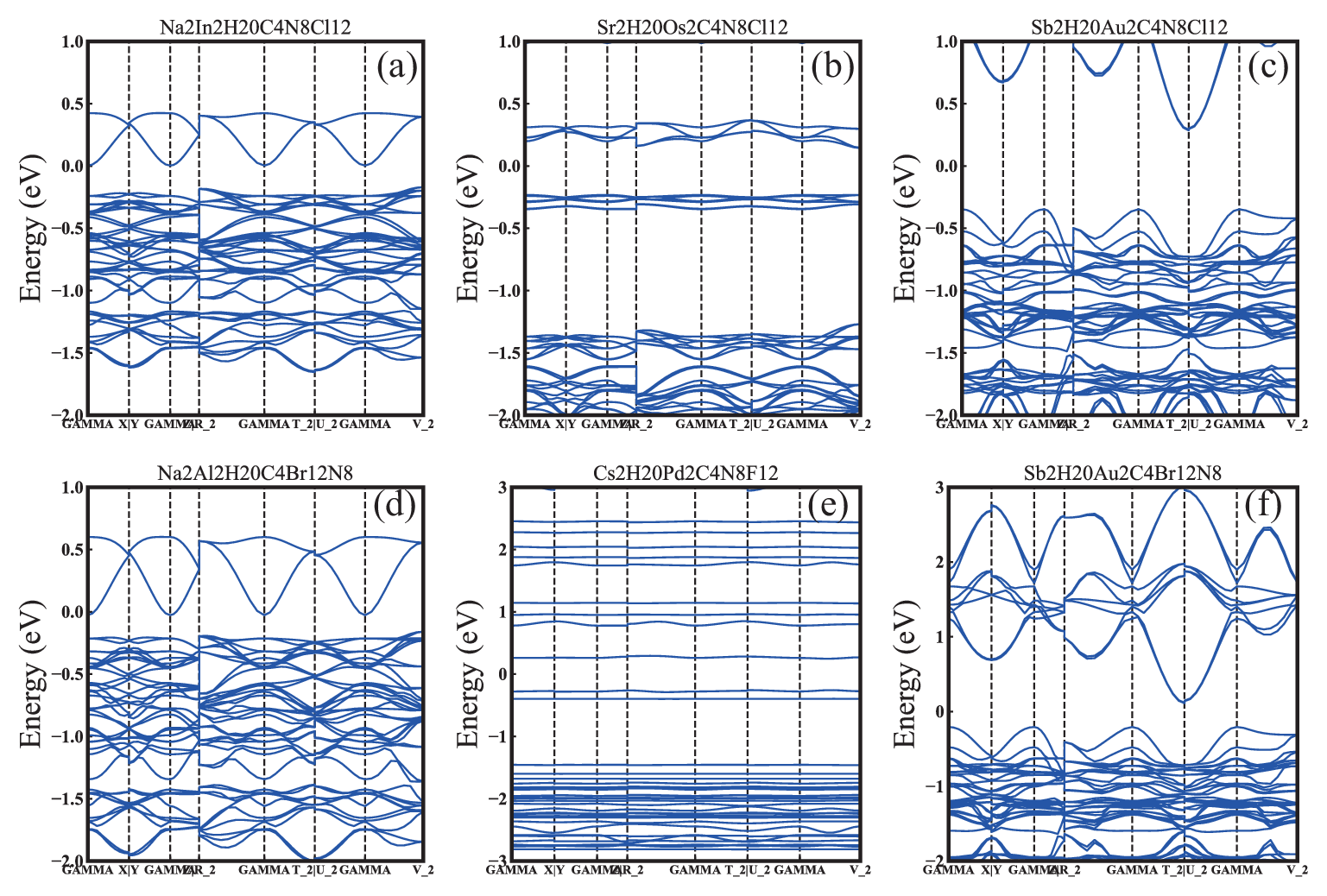}
\caption{Calculated PBE bandstructures of (a) (CH$_3$)$_2$NH$_2$NaInCl$_6$, (b) (CH$_3$)$_2$NH$_2$SrOsCl$_6$, (c) (CH$_3$)$_2$NH$_2$AuSbCl$_6$, (d) (CH$_3$)$_2$NH$_2$NaAlBr$_6$, (e) (CH$_3$)$_2$NH$_2$CsPdF$_6$, and (f) (CH$_3$)$_2$NH$_2$AuSbBr$_6$. }
\label{fig:bands} 
\end{figure}

The six (CH$_3$)$_2$NH$_2$-based HOIDP candidates with finite bandgaps are  (CH$_3$)$_2$NH$_2$NaInCl$_6$, (CH$_3$)$_2$NH$_2$SrOsCl$_6$, (CH$_3$)$_2$NH$_2$AuSbCl$_6$, (CH$_3$)$_2$NH$_2$NaAlBr$_6$, (CH$_3$)$_2$NH$_2$CsPdF$_6$ and (CH$_3$)$_2$NH$_2$AuSbBr$_6$, and the PBE-calculations of their bandstructures are shown in Figure~\ref{fig:bands}(a-f), respectively. The calculated PBE bandgaps are 0.173 eV, 0.376 eV, 0.633 eV, 0.132 eV, 0.504 eV and 0.319 eV, respectively. Since the PBE method always underestimates the real bandgaps, the chosen six  (CH$_3$)$_2$NH$_2$-based HOIDP candidates are reasonably believed to possess real bandgaps around 1.0 eV, which is beneficial for photovoltaic applications.

It should be noted that, limited by our computation resources and the intense computation required for DFT calculations, high-throughput DFT calculations of all filtered HOIDP candidates with different organic groups and appropriate bandgaps are still needed to perform in the following work. Moreover, the calculations of optical absorption and the subsequent photovoltaic parameters are also needed to be conducted to obtain the theoretical photovoltaic performances of our predicted HOIDP candidates finally.


\subsection{Discussions}

We believe that there is still room for improvement in the existing approaches, starting with the fact that the training set for the CGCNN method is too small. We use more than three thousand materials as the training set, but for a neural network of this size, more than three thousand materials do not perform very well. 
~Secondly, when selecting materials, we only consider the material properties, but not the toxicity and rarity of the materials\cite{fab-all,Wu2019}, some materials are too rare or toxic to be used even though they are good in nature. Therefore, we should give more weight to some very abundant materials because they are cheap. If the scale can be increased, then the higher efficiency brought by the material change is not so obvious. Finally, further work should be devoted on the finely tuning of the hyper-parameters of the CGCNN method, or improving the neural network of CGCNN by introducing some new mechanisms, such as the self-attention mechanisms\cite{wang2021compositionally}.

\section{Conclusion}
In our research, we analyze features of importance of high precision classic machine learning models, and we conclude some features that high-performance perovskites generally have, and we also analyze composition and conclude some trends on molecules that compose A-site, B-site, and X-site in perovskites. For example, atom which is more reactive in X-site of perovskite more likely results in a bandgap not within 0.5 eV and 2.0 eV. Then, We find the potential of CGCNN compared to classical machine learning because of the information obtained by acquiring more structure, achieving close to classical machine learning with fewer samples and very few features. Because of the better representation of molecular structure features, CGCNN has better interpretability. We also combine classic machine learning and CGCNN together to screen materials. These materials might be higher-performance Perovskites in Solar Cells to alleviate energy shortage. We believe that if more structure files can be integrated as input to CGCNN, CGCNN will have great potential to surpass classical machine learning and predict higher-performance perovskites to optimize the efficiency of solar cells. In the end, we find out that double perovskites based on (CH$_3$)$_2$(CH$_2$)$_2$NH$_2$, CH$_3$(CH$_2$)$_3$NH$_3$, (CH$_3$)$_2$NH$_2$, and C(CH)$_5$NH$_3$ are more likely high-performance perovskites on Solar Cells. In addition, double perovskites with higher reactive halogen on X-site, with Be on B$_1$-site, and with Pt and Pd on B$_2$-site will have better performance in general. We also conclude 6 high performances (CH$_3$)$_2$NH$_2$-based HOIDPs-(CH$_3$)$_2$NH$_2$NaInCl$_6$, (CH$_3$)$_2$NH$_2$SrOsCl$_6$, (CH$_3$)$_2$NH$_2$AuSbCl$_6$, (CH$_3$)$_2$NH$_2$NaAlBr$_6$, (CH$_3$)$_2$NH$_2$CsPdF$_6$ and (CH$_3$)$_2$NH$_2$AuSbBr$_6$-by combination of classic machine learning and CGCNN.

\newpage
\section*{References}
\bibliographystyle{plain}	
\bibliographystyle{unsrt}
\bibliography{main}

\newpage
\section*{Acknowledgement and Authors Contributions}

Xinjian Qiu helped launch and design this project. Xinjian Qiu, Linkang Zhan and Danfeng Ye together selected the topic, and Xinjian Qiu performed the research of background. In this thesis, firstly, Linkang Zhan did most of the thesis writing, classical machine learning modeling, and modeling with graph neural network models, and plotted the results. Danfeng Ye's main work is to perform the ab-initio calculation, which is used to verify the accuracy of the results calculated by graph neural networks and classical machine learning. Xinjian Qiu also wrote some of the thesis. We are grateful to our advisor Yan Cen for leading us without paying to understand the use of graph neural networks in materials, teaching us the background knowledge about HOIDPs and the principles and methods of ab-initio calculation. All authors analyzed and discussed the results. 

\end{document}